**Title**: An Iterative, User-Centered Design of a Clinical Decision Support System for Critical Care Assessments: Co-Design Sessions with ICU Clinical Providers

**Authors**: Andrea E. Davidson[1,2], Jessica M. Ray[3], Ayush K. Patel[1,2], Yulia Strekalova Levites[1,4], Parisa Rashidi[1,5], Azra Bihorac[1,2]

[1]Intelligent Clinical Care Center, University of Florida, Gainesville, Florida, USA

[2]Division of Nephrology, Hypertension, and Renal Transplantation, Department of Medicine, College of Medicine, University of Florida, Gainesville, Florida, USA

[3] Department of Health Outcomes & Biomedical Informatics, College of Medicine, University of Florida, Gainesville, Florida, USA

[4]College of Public Health and Human Professions, University of Florida, Gainesville, Florida, USA

[5]J. Crayton Pruitt Family Department of Biomedical Engineering, University of Florida, Gainesville, FL, USA

**Abstract** (not to exceed 450 words)

Introduction: This study reports the findings of qualitative interview sessions conducted with ICU clinicians for the co-design of a system user interface of an artificial intelligence (AI)-driven clinical decision support (CDS) system. This system integrates medical record data with wearable sensor, video, and environmental data into a real-time dynamic model that quantifies patients' risk of clinical decompensation and risk of developing delirium, providing actionable alerts to augment clinical decision-making in the ICU setting.

Methods: Co-design sessions were conducted as semi-structured focus groups and interviews with ICU clinicians, including physicians, mid-level practitioners, and nurses, at a 1,000+ bed tertiary hospital in the Southeastern United States. Study participants were asked about their perceptions on AI-CDS systems, their system preferences, and were asked to provide feedback on the current user interface prototype. Session transcripts were qualitatively analyzed to identify key themes related to system utility, interface design features, alert preferences, and implementation considerations.

Results: Ten clinicians participated in eight sessions. The analysis identified five themes: (1) AI's computational utility, (2) workflow optimization, (3) effects on patient care, (4) technical considerations, and (5) implementation considerations. Clinicians valued the CDS system's multi-modal continuous monitoring and AI's capacity to process large volumes of data in real-time to identify patient risk factors and suggest action items. Participants underscored the system's unique value in detecting delirium and promoting non-pharmacological delirium prevention measures. The actionability and intuitive interpretation of the presented information was emphasized. The participants also noted that system alerts should role-specific and that thresholds should be carefully established to reduce alert fatigue.

Conclusion: ICU clinicians recognize the potential of an AI-driven CDS system for ICU delirium and acuity to improve patient outcomes and clinical workflows. Successful adoption will depend on effective integration into clinical practice, an intuitive design, and careful management of alerts. Future work should further address implementation considerations and test the usability of the revised interface prototype.

## Introduction

Modern intensive care units (ICUs) are dynamic environments where patient status can shift rapidly, requiring attentive monitoring and swift intervention [1]. These critical care patients are often faced with multi-organ system conditions that require life-sustaining interventions such as mechanical ventilation, vasopressor medications, blood transfusions, and continuous renal replacement therapy [2], [3]. Such interventions require precise recognition of changes in patient status to guide effective clinical decisions. Furthermore, the ICU environment itself—marked by constant noise, sleep disturbances, prolonged immobilization, use of sedation, and mechanical ventilation—create conditions that contribute to cognitive impairment and delirium [4]. Delirium is an acute neurocognitive condition characterized by disturbances in attention, awareness, and the ability to orient oneself to the environment [5], [6], [7]. It represents a significant challenge in critical care medicine, affecting up to affects up to 61% of adult ICU patients, and is associated with longer hospital stays, increased mortality, and cognitive impairments [6], [7]. Despite its high prevalence, delirium is often underdiagnosed and has not shown significant success with pharmacological treatments, underscoring the importance of guided preventative and non-pharmacological interventions [8], [9]. Current approaches to delirium recognition rely on tools such as the Confusion Assessment Method for the ICU (CAM-ICU) and the Intensive Care Delirium Screening Checklist (ICDSC), both of which depend on regular assessments and clinician training [10]. Treatments focus on addressing modifiable risk factors, such as minimizing sedation, promoting early mobilization, and improving sleep hygiene, rather than relying on pharmacological interventions, which have shown limited efficacy [4].

Timely identification of cognitive changes and acuity levels is crucial for successful patient outcomes by enabling early interventions and reducing complications [11]. However, traditional patient monitoring methods are limited by the availability of and processing capacity of ICU nurses and physicians. Bedside nurses manage large volumes of real-time data, such as vital signs and laboratory results, while physicians and mid-level providers interpret this information and make decisions. With some ICU providers managing over 20 patients, the continuous data influx creates a bottleneck in care, leading to cognitive biases and inconsistencies in patient management [1], [12]. Clinician assessments can also be affected by experience level, making it harder to detect subtle signs of delirium or cognitive decline [13]. This dependence on clinician presence and manual interpretation can result in delays and missed identification of subtle but meaningful changes in patient status. These overwhelming demands make individualized tracking of patient trajectories challenging and inconsistent, ultimately affecting the quality of care [14], [15].

In the critical care setting, rule-based scoring systems such as the Sequential Organ Failure Assessment (SOFA) and Modified Early Warning Score (MEWS) aid in risk classification but are static and depend on manual input [16]. Artificial Intelligence (AI)-enhanced decision support systems have emerged as Clinical Decision Support (CDS) systems, leveraging powerful machine learning (ML) algorithms to analyze large datasets and identify subtle patterns that may indicate an impending change in condition. AI/ML models are trained on vast cohorts of patient data, enabling them to learn from complex patterns and apply those insights in real-time. This ability to continually process and interpret data to provide actionable recommendations distinguishes AI-enhanced tools from traditional, static monitoring systems, making them invaluable in high-acuity environments such as the ICU [16], [17]. To support these predictive capabilities, AI-CDS system leverage advanced sensing technologies to integrate multimodal data sources enabling a continuous flow of real-time data to provide a comprehensive assessment of patient status. By enhancing clinical insights and supporting proactive decision-making, this approach addresses critical gaps in traditional monitoring practices [18], [19].

For CDS systems to be effective in clinical settings, they must align with local workflows and user needs. Hence, user-centered design (UCD) and co-design methodologies are key for creating systems that integrate seamlessly into clinical practice. UCD involves end-users in the design process to ensure tools meet real-world needs, while co-design fosters collaboration between developers and users throughout iterative development [19], [20]. These approaches are essential for AI-based CDSS in critical care, where tools must balance advanced functionality with ease of use, reliability, and efficiency while preserving clinical autonomy [20]. This work applies UCD and co-design to an autonomous AI-CDS system that integrates sensor-based monitoring to overcome the limitations of traditional systems that rely on static data and often miss small, clinically significant changes. This combination enhances adoption and effectiveness to address the practical needs of ICU providers as end-users, paving the way for more practical, actionable solutions in critical care [21], [22].

## **Methods**

*Study Context*

This study was conducted as part of a larger project to develop an autonomous sensing system that integrates medical record and data from wearable devices, video and depth cameras, and environmental sensors into a real-time dynamic model for the identification of acuity and delirium risk status of patients receiving care in ICUs. The results from the AI algorithms are displayed to clinical providers within a user interface presenting the calculated acuity and delirium risk scores, the top contributing risk factors to each score, an interactive graph, and sensor data summaries. Moreover, this system acts as an early warning system, delivering actionable prompts to users when risk scores surpass a set threshold.

We applied a qualitative co-design methodology to the design of the user interface of this CDS system engaging ICU providers throughout an iterative design process. The present study reports the findings from the first phase of this co-design process and seeks to understand clinicians' perceptions of AI-supported CDS systems, elicit feedback on user interface and system features, and identify the unique needs of ICU providers.

*Study Design*

Co-design sessions were structured as semi-structured focus groups or individual interviews and took place in-person and virtually via Zoom meetings. Each session was audio recorded and transcribed using a two-person approach that included redaction of personally identifiable information. Participants also completed a brief demographics survey. This study was reviewed and approved by the University of Florida (UF) Institutional Review Board prior to its implementation.

The interview guide was written by a faculty member from the UF College of Public Health and Human Professions and can be found in the Supplemental Materials. Following a brief introduction on the autonomous sensing and CDS system, participants were asked questions covering five general topics: (1) the provider's mental model of autonomous sensing systems and their role in clinical decision-making, (2) system utility, (3) technical feature preferences and alert sensitivity, (4) interface design, and (5) implementation considerations. During the sessions, participants were shown images of the preliminary interface prototype to probe for feedback and discussion on system features, visual design elements, and prompt sensitivity. These interface mock-ups are located in the Supplemental Materials.

*Participants*

Eligible participants were adult clinical providers— including physicians, physician assistants, nurse practitioners, and registered nurses—who work in any intensive care unit at a 1000+ bed tertiary hospital in the Southeastern United States. These individuals were identified using institutional directories and listservs, with recruitment conducted through direct solicitation via email. Purposeful sampling was employed to ensure the inclusion of participants with diverse roles within the clinical setting. Informed consent was obtained from participants at the beginning of each session.

*Analysis*

A structured codebook was created (Supplemental Materials 3), and qualitative coding of the session transcripts was performed by a two-member team. In the first round of coding, the team independently applied the codebook to the transcripts, identifying relevant codes and recurring patterns. These codes were then grouped and synthesized into broader themes and subthemes based on the recurrent ideas in session transcripts. A second round of coding was conducted, during which the team further refined the themes and subthemes through iterative discussions.

## **Results**

Study Participant Characteristics

A total of 10 clinicians participated across 8 co-design sessions—1 focus group of 3 participants and 7 individual interviews. Those interviewed were primarily white non-Hispanic (N=9) and included 1 nurse, 2 physician assistants, 1 resident physician, 1 Fellow physician, and 4 attending physicians. Participant characteristics can be found in Table 1. Most participants were 36-40 years old (N=4), followed by 31-35 years old (N=3), 25-40 years old (N=2), and 41-45 years old (N=1).

Table 1. Participant Demographics

| Demographic | | Percent | N |
|---|---|---|---|
| **Gender Identity** | Man | 30% | 3 |
| | Woman | 70% | 7 |

| | Race and Ethnicity | | | |
|---|---|---|---|---|
| | | Asian/Pacific Islander | 10% | 1 |
| | | Hispanic/Latino | 0% | 0 |
| | | White | 90% | 9 |
| | **Role** | | | |
| | | Attending Physician | 40% | 4 |
| | | Resident/Fellow Physician | 30% | 3 |
| | | Physician Assistant (PA) | 20% | 2 |
| | | Registered Nurse (RN) | 10% | 1 |
| | **Setting of Practice** | | | |
| | | Neuromedicine ICU | 30% | 3 |
| | | Surgical ICU | 40% | 4 |
| | | Cardiac ICU | 20% | 2 |
| | | Anesthesiology | 10% | 1 |
| | **Age** | | | |
| | | 25-30 | 20% | 2 |
| | | 31-35 | 30% | 3 |
| | | 36-40 | 40% | 4 |
| | | 41-45 | 10% | 1 |

Qualitative Results

Qualitative analysis of interview transcripts identified five themes characterizing the perceptions of ICU providers on the AI CDSS: (1) AI's computational utility, (2) workflow optimization, (3) effects on patient care, (4) technical considerations, and (5) implementation considerations. A summary of these themes with their subthemes and definitions are presented in Table 2. Key quotes from each subtheme are found in Table 3.

Table 2. Taxonomy of Themes from Clinician Discussions of an AI Pervasive Sensing System

| Theme | Subtheme | Definition |
|---|---|---|
| AI's Computational Utility | Early warning system | AI's ability to detect patterns and trends early to identify potential patient risks before they become clinically apparent. |
| | Data processing and visualization | AI's capability to process large datasets and present them visually, helping clinicians identify trends and make data-driven decisions, including personalized predictions based on patient history. |
| Workflow Optimization | Personal-level optimization | The system's ability to optimize personal workflows and mitigate information overload by identifying high-risk patients, summarizing key data, and providing recommendations, especially for less experienced staff. |
| | Team-level optimization | The system's effect on the collective ICU work environment by encouraging communication, introducing new workflows and action items, and increasing compliance with preventative delirium measures. |
| Effects on Patient Care | Continuous pervasive monitoring | Continuous multi-modal data collection and real-time monitoring to collect data outside the standard of care and catch important events or changes that might be missed between regular assessments. |
| | Improvements to delirium management | The system's potential to standardize delirium detection, provide earlier interventions, and maximize the implementation of non-pharmacological preventative measures. |
| | Better assessment for patients with communication difficulties | The system's ability to detect signs of pain and delirium in patients who cannot communicate verbally by analyzing facial expressions, mobility data, and physiological data. |
| Technical Considerations | Alert preferences | The preferences on system alerts: selecting the appropriate individual to send each alert, sensitivity vs. specificity, and the appropriate number of daily alerts. |
| | Design features | Participants' desired design features, suggestions for a user-friendly design, clear data visualizations, and interactive features that enhance the system's usability. |
| | Platform preferences | The importance of integrating the AI system with existing medical record platforms like Epic and its availability on web and app platforms. |
| Implementation Considerations | System adoption | Strategies for promoting system adoption, including integration into hospital policies and workflows, and providing sufficient training and support. |
| | Future unknowns | Uncertainties regarding the system's implementation, including impact on staff workload and potential ethical and legal concerns. |

*AI's Computational Utility*

The system's ability to serve as an early warning system, identifying subtle trends and patterns before they become clinically apparent, was seen by participants as its most valuable feature. Participants emphasized that early recognition of changes in patient status could allow for proactive interventions and potentially improve patient outcomes. However, responses varied regarding the desired level of guidance. Some participants expressed a preference for receiving actionable recommendations and clinical guidelines, while others, particularly more experienced providers, preferred to create their own action items based on the top contributing factors identified by the model. This variation underscores the need for actionability while preserving clinician autonomy.

Participants also acknowledged that AI's ability to process vast amounts of data goes beyond human capacity, allowing it to recognize patterns that might otherwise be overlooked. As one participant described, "sometimes you don't realize what's actually happening until you're that person that comes on service after being off for a couple weeks and then you're like, 'Oh I think there's something wrong here, there's something we're missing,' but you're caught up in the weeds that you don't see that pattern because you're living it every single day with that patient." User exploration of the trended data was also discussed, and the ability to visualize processed data in an interactive platform was seen as uniquely beneficial, "It would be nice to be able to maybe click on the icons that are available in each category […] So that you have the option to get more data if you want, and then again you can look at trends within each category."

Participants emphasized the need for a system that provides both a high-level summary of patient status, using graphs, trends, and delta changes, as well as the ability to drill down into more granular data when necessary. Additionally, while not a current feature of our models, they discussed the potential benefit of a system that could integrate patient history into predictive modeling. If a patient had a prior history of delirium, for instance, the system could adjust its risk prediction accordingly and suggest interventions that had previously been effective, thus promoting personalized and preventative care.

*Workflow Optimization*

Participants further described how the system could potentially fit into their workflow and improve efficiency. Clinicians noted that the ICU environment's high cognitive load and learning curve, and the overwhelming amount of data contribute to decision fatigue and burnout. Many participants welcomed the system as a tool that could intelligently process data, present interpretable findings, and mitigate information overload. Several also noted that the system's risk stratification capabilities could help junior clinicians develop an understanding of key clinical parameters and warning signs. Additionally, the system was viewed as a valuable tool for shift transitions, helping clinicians quickly orient to their assigned patients' hospital course when coming on service and improving continuity of care between day and night shifts.

One recurring discussion was this system's ability to catalyze team-based care, encouraging dialogue and discussion on changes in patient status or risk factors. Many cited delirium as being overlooked, but that this system's integration into the workflow could make delirium precautions a more consistent point of discussion. The use of this system to measure compliance was also discussed, with one participant saying, "various things will be ordered for the patient that are not necessarily implemented reliably. And I think that using an AI system to help monitor the adherence to the delirium precaution that you're implementing"

Beyond improving communication, some noted that risk scores could allocate resources more effectively, allowing clinicians to prioritize interventions or involve additional team members earlier. Others noted that the system could encourage support staff involvement, particularly if nonpharmacological interventions were tracked and linked to improvements in patient scores. As one phrased it, "I feel like those things will matter more when we have data to support that—look, all the noise was loud, they were uncomfortable, they were woken up, all these things that are right in our face saying, 'they are going to start becoming delirious the next day if you don't do something about it.'"

Similarly, another significant area of discussion was the workflow or cascades that this system would create. Participants imagined that this would be similar to current practices where patient abnormalities are monitored, communicated, and escalated up a chain of command. Under this workflow, nurses generate clinical scores and document assessments, communicating with residents and attendings when concerning values arise. Some participants mentioned that these scores could function like the Modified Early Warning Score (MEWS), where nurses are required to notify a physician if the value surpasses a certain score. Several physician participants discussed how they take things more seriously when a nurse comes to them with concern. One physician assistant explained, "you're looking at a lot of numbers to get lost or to forget about, or kind of push it to the back because you're getting so many updated chart alerts. But when a nurse comes and tells me something, I think that triggers something in my brain that okay they're actually really concerned so I should probably like think about this twice." However, participants cautioned that obligatory actions due to elevated risk scores could prompt clinicians to make decisions they wouldn't otherwise make, inadvertently leading to unnecessary testing or overtreatment.

*Effect on Patient Care*

Participants valued the system's ability to capture data continuously outside of the standard of care (e.g., video and environmental data), noting that this could identify clinically significant events that might otherwise be missed between routine check-ins. Some described the system as a "second set of eyes" for busy ICU settings. While a few participants speculated that continuous AI-driven monitoring might reduce some staffing burdens, most focused on its utility in catching something that could have gone overlooked. As one attending put it, "medical errors are not that someone makes a bad decision most of the time. Most of the time it's more of the prioritizing, or paying attention to things that get missed."

Delirium management was an area identified where continuous monitoring would address an unmet need, with one participant saying, "I think that within our ICU rounding round we're not really good at doing our CAM scores daily and checking for delirium. So, I think this will be helpful, doing more of a continuous measurement, since it can be in the room much more frequently than our bedside folks can." Delirium was a primary focus of discussion, with participants highlighting its diagnostic challenges and similarities to other neurological issues, the experience required for recognition, and the limited availability of pharmacologic treatments. The system's ability to provide continuous monitoring for subtle cognitive changes was seen as an advancement in both prevention and early intervention. "All of those things would be great to alert a clinician who may not necessarily be paying attention to that […] so that you can implement delirium prevention measures earlier and perhaps achieve a more beneficial outcome," one surgical resident summarized. Additionally, the potential for AI to phenotypically classify delirium subtypes could enhance diagnostic accuracy and facilitate more tailored management approaches.

Participants identified the system's ability to autonomously assess pain and agitation as a promising feature for nonverbal or intubated patients. The system's capacity to analyze facial expressions in conjunction with physiological data could improve the recognition of pain in sedated or hypoactive delirious patients, leading to more timely and appropriate interventions.

*Technical Considerations*

In discussing integration into the clinical setting, all participants strongly preferred integration into existing electronic health record (EHR) systems, Epic, which has a web- and app-based platform. Lack of integration was viewed as a significant barrier to adoption, as clinicians expressed concerns about workflow disruptions and the risk of forgetting to check an independent system. Physicians who see patients across several ICUs noted that they do not always have access to a computer. This mainly affects the model's alerts and prompts rather than checking the risk scores or data visualization features, which clinicians generally said they would only use once daily.

Participants expressed concerns about alert fatigue and emphasized the need for role-specific notifications. The consensus was that bedside nurses should receive alerts related to actionable interventions (e.g., environmental modifications), while attending physicians preferred to receive notifications only for significant changes in patient status requiring immediate attention. When speaking about alerts on environmental determinants of delirium, one attending physician said, "I don't want to be alerted. I want that data to be available to me on route, and more importantly, I think I would like that to go to the patient's bedside nurse. I think that's what's makes a difference, because that's the person who's going to make those changes." Delta changes in risk scores were also highlighted as particularly important; clinicians noted every patient receiving care in an ICU is a high-risk patient, so the system should focus on identifying meaningful shifts and underlying factors rather than static risk levels.

Alert thresholds were discussed and the importance of finding an appropriate threshold was emphasized to limit alert fatigue. Aside from attending physicians, one alert per day (or per shift) was typically deemed appropriate, but several participants preferred to err on the side of caution and receive more false positives.

The design preferences participants shared emphasized quick interpretability and ease of use. The thresholds defining low- and high-risk scores should be clearly defined, and the risk factors should be displayed to explain why the patient has been identified as such. Attention to the meaning behind the colors used to display risk scores must be maintained. Participants emphasized data visualization as a critical feature of the design interface, requesting the ability to hover over data points on the graph and ability to expand for more detailed data. The ability to see what times certain events are happening—such as pain, mobility, or noise—were important to allow clinicians to consider the context behind such occurrences.

*Implementation Considerations*

Participants agreed that widespread adoption of the system would be most effectively achieved through hospital policy mandates. However, they emphasized that ease of use and demonstrated patient benefit would be key factors in voluntary adoption. Implementation strategies such as pilot programs, in-person training and workshops, and self-paced online modules were suggested as methods for increasing engagement.

Several implementation challenges were identified, including the feasibility of installing the system in every patient room and the potential for increased workload for bedside nursing staff. Participants also raised concerns about the downstream effects of risk scoring, particularly regarding its influence on discharge decisions, ICU transfers, and post-acute care placement. The potential for legal implications—such as liability in cases where alerts were muted or ignored—was also identified.

Table 3. Key Quotes Across Themes

| Theme | Subtheme | Quotes |
|---|---|---|
| AI's Computational Utility | Early warning system | "I feel like that's where AI/machine learning can help synthesize some of that pattern for us and show us blood pressure's going up, heart rate is going down, they are herniating, and I just can't see it quite yet because it's happening very slowly." *(R1)* |
| | Data processing and visualization | "[…] the biggest goal, I think, in terms of using AI would be to identify trends in that continuous data all day long, and have like built-in alerts to let us know when it notices a trend that we've pre-specified as being concerning so that we don't miss anything, because like I said a lot of this variation happens second-to-second, so just because that value may have sort of gone up for a second back up to a more normal value, and that's the one that gets charted, and then kind of goes down. You could be missing something." *(R3)* |
| Workflow Optimization | Personal-level optimization | "[…] it's hard to take all that data, and then funnel it into something that's useful, and [this system] helps you triage which person you go check on first.*(A3)* |
| | | "And when you're kind of in the thick of it sometimes you don't realize what's actually happening until you're that person that comes on service after being off for a couple weeks and then you're like, 'Oh I think there's something wrong here, there's something we're missing,' but you're caught up in the weeds that you don't see that pattern because you're living it every single day with that patient." *(R1)* |
| | Team-level optimization | "I think the best way I would agree is to have it similar to sepsis when the nurse is alerted they have to come get you, because I think it encourages the dialogue between you and your provider team. […] it provides conversation about an objective number that the nurse now has to watch for" *(P1)* |
| Effects on Patient Care | Continuous pervasive monitoring | "Definitely, because it's in the room consistently. I think it will catch things during the day as far as hyperactive delirium. If you guys are monitoring faces like that, if somebody is consistently in pain, I think that's going to be extremely helpful, since I could contribute to a patient's delirium in the ICU environment. I think, having that constant supervision will be helpful." *(A1)* |
| | | "[…] provides the opportunity to help with staffing stretches when we unfortunately have to increase our patient ratio to staff, if they are good at telling us things are going awry, […] that may unburden the stress of the caregivers, knowing that there's sort of a backup system if they end up having to be trapped in another room taking care of somebody for longer than they had anticipated." *(A2)* |
| | Improvements to delirium management | "It's really easy to let these things go because it becomes sort of like everyone talks about delirium precautions, but it's not always practiced as such. But if we know this patient is elderly, they're at high risk, then on rounds we have to talk about polypharmacy, we have to think about these subtle things like asking the family members to come help us keep them awake during the daytime. I feel like those things will matter more when we have data to support that— look all the noise was loud, they were uncomfortable, they were woken |
| | | up, all these things that are right in our face saying, they are going to start becoming delirious the next day if you don't do something about it." *(R1)* |
| | Better assessment for patients with communication difficulties | "It would be really to recognize delirium early before I had a chance to recognize it, and to try to look into delirium patients who actually can't express themselves very well. For example, somebody who's intubated, and I'm trying to extubate them. It will be nice to know why they are delirious." *(A4)* |
| | | "I am interested in pain responses especially for our intubated patients, who we have a very difficult time assessing pain scores on. So, if it's watching the face and it's sees that the patient has been grimacing for extended periods of time or is always having pain issues earlier than their scheduled analgesics, that would be helpful." *(A2)* |
| Technical Considerations | Alert preferences | "[…] if I were to get a single alert per day for every patient, that would be too much. It would just add to the burden of filtering through all the alerts. It'd be too hard. If I take myself back to being like a resident and I've got 10 patients in the ICU, maybe 1 or 2 per person, so 10 to 20 is still a lot." *(A3)* |
| | Design features | "It would be nice to be able to maybe click on the icons that are available in each category so that you can see, for example, if you're really interested in an itemized number of disruptions or duration, or you could just get like an overall duration of disruptions or the evening. So that you have the option to get more data if you want, and then again you can look at trends within each category." *(R2)* |
| | Integration into the current electronic medical record system | "[…] having an Epic app is pretty simple or using something within Epic is pretty simple, because now that we've gone to the chat feature, I think that's been extremely helpful in improving communication because I can just pull stuff up on my phone and I don't actually have to log into a workstation to get messages that were sent to me." *(A1)* |
| | | "I think it has to be integrated. Yeah, I mean, if it's to be used. I think it will then be integrated into Epic." *(A4)* |
| Implementation Considerations | System adoption | "A big component is, make sure that it integrates with the practice that already exists, as opposed to trying to revolutionize the practice" *(A3)* |
| | Future unknowns | "[…] if someone is being dispositioned to home who has a high-risk score, then what are the implications if they have an adverse event at home, and then family is seeking legal recourse, and they utilize this available dispositional piece of data or information as a justification for why malpractice ensued in their minds. […] Or what if the facility starts using this information and rejects patients for a certain level of care?" *(R2)* |

Table Legend: Participants have been assigned an anonymous identification code according to their role in the clinical setting. Nurse (*N1*), physician assistants (*P1-P2*), resident and fellow physicians (*R1-R3*), and attending physicians (*A1-A4*).

## Discussion

This study highlights key ways ICU clinicians envision AI augmenting their workflow and decision-making, particularly in the setting of delirium recognition and management. Overall, clinicians viewed our AI-CDSS supported by autonomous sensor-based monitoring technologies as a potentially valuable support system for early identification of patients at risk for delirium in the ICU. The primary value of the system lies not in the risk score itself, but in its ability to synthesize complex data into clinically meaningful predictions. Rather than replacing clinician judgment, it serves as a decision aid that enhances situational awareness and promotes evidence-based communication and intervention. Our findings further suggest, however, that system adoption will depend on its integration into the current workflow, proven ability to improve outcomes, and present interpretable findings beyond human computational capacity.

Our findings advance the literature of AI-enabled CDSS by capturing clinician perceptions of key features for meaningful integration into clinical care environments. In particular, the use of sensor-based monitoring to advance beyond the traditional rule-based CDSS or even the retrospective data models of many current AI systems, sets the current work apart as an important advancement in change detection and early warning systems. Traditional early warning systems have used observational data updated by hand while our system offers, as our participants highlighted, an opportunity capture data beyond the current standard of care [23]. Many of our findings also align with prior research on CDS systems and AI-driven clinical decision support, reinforcing the necessity of human-centered AI design [24]. In particular, the extensive discussion on collective and social implications highlight Sendak et al.'s approach of treating AI-CDS systems as "socio-technical systems" to emphasize the interconnectedness of technology and social dimensions [25]. Under this perspective, the end-users' beliefs, contexts, and power hierarchies from technical and institutional infrastructure shape the development and use of AI-CDS systems [25]. Moreover, the participants' recommendations for integration into existing EHR-based workflows reinforce the need for understanding and building AI-CDSS that fit within the local context. Adopting common warning features such as standard use of warning colors is broadly recommended and reinforced within our findings [26], [27]. Finally, a common concern with the introduction of any new detection system is adding to the already overwhelming amount of information and alerts within the clinical setting. Our findings offered initial guidance and a reinforcement of the importance of providing the right alerts and information to the right individuals at the right time, in line with what is commonly known as the five rights of CDS [28].

This study leveraged a co-design approach, ensuring that system features we develop will directly address real-world clinical needs. The inclusion of diverse roles—nurses, midlevel providers, residents, and attending physicians—allowed for a comprehensive assessment of interdisciplinary and chain-of-command implications. However, this is a single-site study, and our findings are limited by the small number of participants at the single institution. We were, however, able to recruit from across multiple ICUs within our institution providing perspectives from a range of clinical experiences and patient populations treated. Our sample was also potentially subject to self-selection bias, as it is possible that we only received interest from recruitment emails from individuals already interested in AI healthcare applications. In fact, multiple participants shared that they were either interested in or already involved in AI research endeavors. Additionally, the study was conducted in an academic institution with significant AI research investment, which may not reflect broader clinical settings though this is likely rapidly changing as AI becomes increasingly adopted in both clinical care and everyday life.

Our qualitative findings brought up several topics that should be further investigated and considered during implementation periods. Several critical questions remain unanswered, including ethical considerations surrounding alert muting, liability for missed alerts, and downstream implications of risk scoring on discharge planning. Our future work in the iterative design process includes refining the user interface and conducting additional co-design sessions to gather additional design feedback. During these sessions, a "think aloud" technique will be used to optimize design preferences, usability features, and interpretability of model findings.

**Conclusion**

This study underscores the potential of AI-driven decision support to enhance ICU workflow efficiency, patient monitoring, and interdisciplinary communication. By leveraging AI's computational capabilities while maintaining clinician autonomy, this system offers a promising tool for improving patient outcomes. However, careful implementation and ongoing evaluation will be necessary to ensure that the system enhances—rather than disrupts—clinical practice. Future work should prioritize user-centered refinements and address outstanding ethical and logistical considerations to optimize real-world applicability.

Supplementary Materials

Title: An Iterative, User-Centered Design of a Clinical Decision Support System for Critical Care Assessments: Co-Design Sessions with ICU Clinical Providers

Supplementary Materials 1: Interview Guide

Hello, my name is __________.

Thank you for agreeing to participate in today's focus group. As you have already heard, we are interested in your thoughts on [insert name they are most likely to have heard this initiative called, maybe ADAPT or I2CU]. Your input will be used to design design displays, set the number of daily alerts, and get the right balance between sensitivity and specificity. To facilitate your input, I will be asking you 9 questions if time permits. There are no right or wrong answers, so answer in any way you wish. You do not have to answer every question, especially if someone has already said what you were going to say. However, we do want to hear from each one of you. And especially if you have a different view. Time is short and there is a lot to cover, so let's keep answers focused. We will be recording, so we don't miss anything, but your answers will not be linked with your personal identity in any way. We will introduce ourselves, but briefly, so we have time for everything. I will start by sharing my name, position at UF and position for this project. I am ______.

[Sensing system shared knowledge/culture; Provider mental model]

1. In general, what role do the autonomous sensing systems, like this one, have in clinical decision-making?

[Usefulness]

2. How can this [system] be useful to you? If so, what is most useful?

[Features: Technical]

3. What features of [the system] are most important to you?

4. How would you like to be alerted about changes in mental status? In environmental determinants of delirium?

5. What number of prompts do you want per patient per day. Let's do a show of hands. 6? 5? 4? 3? 2? 1?

6. Do you prefer high specificity, high sensitivity, or a balance?

[Features: Interface]

7. Here is a mockup of the system interface. What should we keep, what should we drop, what should we have that's not there?

[Implementation: Appropriateness]

8. How could you use the system like this in your clinical practice? What would make this system a good match for your current clinical flow?

    a. Follow up question: Do you want to access the information with an app, the Web, or some other way?

[Implementation: Acceptability]

9. What would make using this system appealing to you?

[Implementation: Feasibility]

10. What would make it easy for you to spend time learning the features of this system or integrating it into your clinical practice?

# Supplementary Materials 2. User Interface Mockups Shown to Study Participants

immobility = lying in bed; mobility =sitting in bed, sitting in the chair, walking with assistance and independent walking

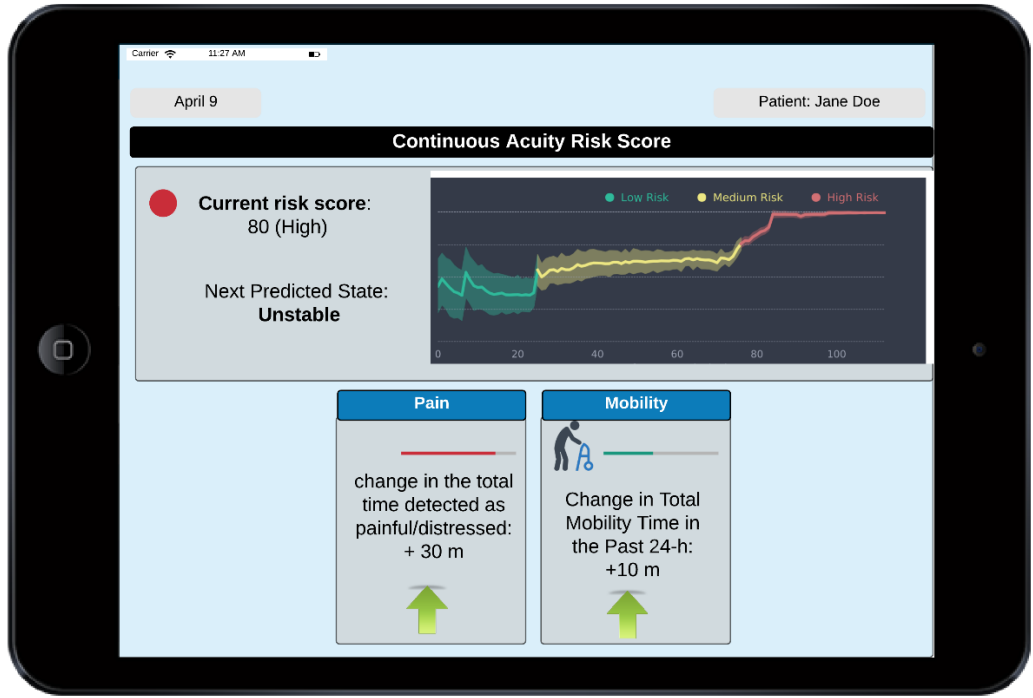

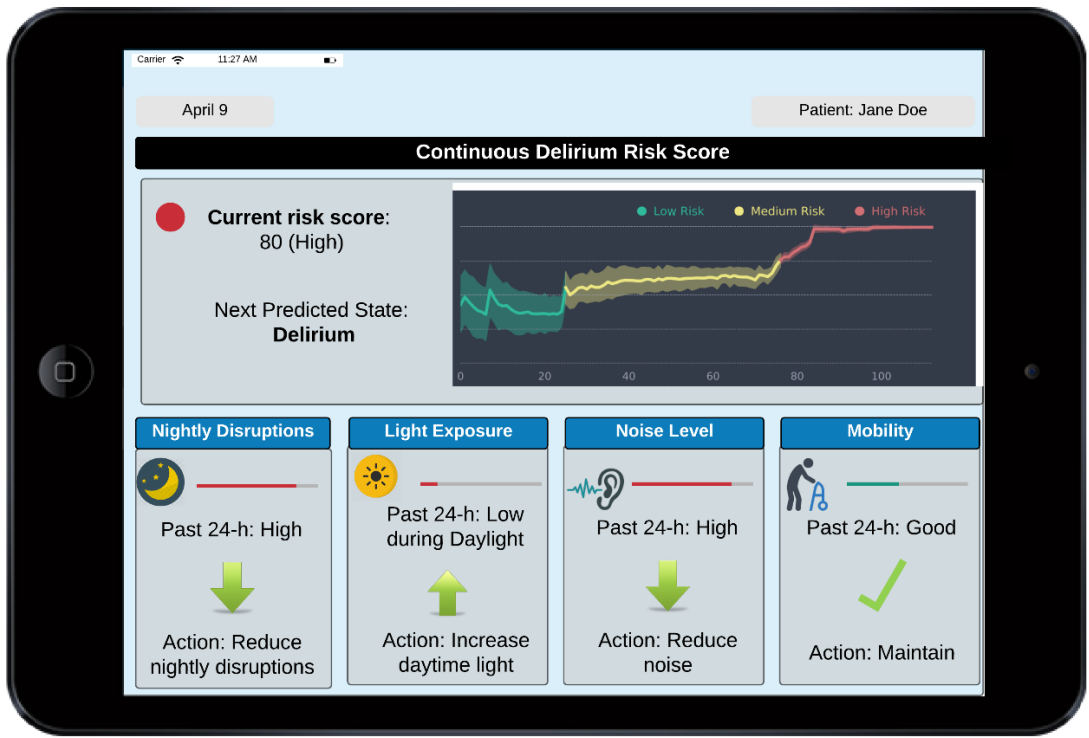

Supplementary Materials 3. Codebook

Coding rules

1. Review the table below and focus on columns Code and Project Definition. Your task is to "code" or assign labels to the sentences of the text that are aligned and representative of the codes below.
2. Before starting to code, read the whole text you are coding quickly. Then, re-read the first sentence and decide if any of the codes from the table below apply.
3. To code, highlight a relevant sentence, add comment, and designate a corresponding code from the Code column of the table below.
4. If a relevant code is identified for the sentence, code all words in the sentence.
5. All relevant sentences (and not just a few examples) from the transcript should be coded. However, not all sentences of each text will be associated with a code. If a sentence does not "fit" any codes, skip it.
6. We will code the texts three times. Once with the codes from Table 1, and then separately with codes from Table 2 and Table 3.

Table 1. Implementation Codes[1]

| Code | Conceptual Definition | Project Definition |
|---|---|---|
| Appropriateness | The perceived fit, relevance, or compatibility of the innovation or evidence-based practice for a given practice setting, provider, or consumer; and/or perceived fit of the innovation to address a particular issue or problem.<br><br>For appropriateness, the criterion is **technical or social**. An EBP can be judged appropriate if it's seen as efficacious for achieving some purpose given existing conditions, including patients' presenting problems, or seen as consistent with norms or values governing people's conduct in particular situations, including organizational mission and treatment philosophy. | Code sentences that describe how the participants could use a system like this in their clinical practice, or what would make this system a good match for their current clinical flow.<br><br>• What makes this system a good match for the ICU setting?<br>• What makes this system align with the goals of ICU providers? |
| Acceptability | The perception among implementation stakeholders that a given treatment, service, practice, or innovation is agreeable, palatable, or satisfactory.<br><br>For acceptability, the criterion is **personal**. Two people can view the same system and form different judgement about its acceptability to them if their needs, preferences, or expectations differ. | Code sentences that describe what makes the AI-supported CDSS an appealing system to the individual.<br><br>• What draws the individual provider to the system?<br>• What makes (or would make) the system appealing to the respondent? |

---

[1] Weiner, B. J., Lewis, C. C., Stanick, C., Powell, B. J., Dorsey, C. N., Clary, A. S., ... & Halko, H. (2017). Psychometric assessment of three newly developed implementation outcome measures. Implementation Science, 12(1), 1-12.

| Feasibility | The extent to which a new treatment, or an innovation, can be successfully used or carried out within a given agency or setting.<br><br>For feasibility, the criterion is **practical**. An EBP can be judged feasible if a task or an action can be performed relatively easily or conveniently given existing resources (e.g., effort, time, and money) and circumstances (e.g., timing or sociopolitical will). | Code sentences that describe what would make it easy for participants to spend time learning the features of the system, or what would make it easier to integrate into current clinical practice. |

Table 2. Technical Feature Codes

| Code | Conceptual Definition | Project Definition |
|---|---|---|
| System Features | A clinical decision support system (CDSS) is intended to improve healthcare delivery by enhancing medical decisions with targeted clinical knowledge, patient information, and other health information.<br><br>The CDSS will function as:<br>1) An adaptive prompting system that will provide prompts for actions aimed to decrease circadian desynchrony and to increase mobility in ICU patients.<br>2) A real-time acuity prediction and visual assessment platform in the ICU environment. | Code sentences where participants speak about what technical features of the system are important to them, their alert preferences, and the sensitivity of the system.<br><br>- How many alerts?<br>- Alert and/or system sensitivity<br>- What the system is sending a notification for |
| Interface features | An interface is a device or a system that unrelated entities use to interact. According to this definition, a remote control is an interface between you and a television set, the English language is an interface between two people, and the protocol of behavior enforced in the military is the interface between people of different ranks.<br><br>Technology and technology-mediated processes are employed by people who need to be capable of navigating them. The end-user will interact with the system's interface—which presents integrated continuous ICU data streamed with deep learning (DL) algorithms. End-user satisfaction and ability to navigate the system interface is the goal of this iterative design process. | Code sentences in which system interface features are discussed, including preferences for how the provider would like to interact with the CDSS.<br>- What platform will be used (web, app, EMR, etc.)<br>- Visual preferences<br>- Functional interface preferences (color symbolism, shapes, etc.) |

Table 3. Agency and Social Codes

| Code | Conceptual Definition | Project Definition |
|---|---|---|
| Self | In agency exercised individually, people bring their influence to bear on what they can control. | Code sentences that describe how the participants plan to use the system themselves. |
| Proxy | In proxy agency, they influence others who have the resources, knowledge, and means to act on their behalf to secure the outcomes they desire. | Code the sentences that talk about what should be and infer that the actions need to be taken by someone else (i.e. nurses, management or administrators, patients' family, institutions, communities, etc.) |
| Collective | In the exercise of collective agency, people pool their knowledge, skills, and resources and act in concert to shape their future. | Code sentences in which participants talk about taking action as part of the group or working with others to take action. |